\documentclass[aip, jmp, reprint]{revtex4-1}

\usepackage{amsmath}
\usepackage{amssymb}
\usepackage{bm}% bold math
\usepackage{dcolumn}% Align table columns on decimal point
\usepackage{graphicx}% Include figure files
\usepackage[mathlines]{lineno}% Enable numbering of text and display math
\usepackage{makecell}  % line breaks in cells

\newcommand{\boltzmann}{k_{\scriptscriptstyle\mathrm{B}}}  % Boltzmann constant
\newcommand{\del}{\partial}  % partial derivative
\newcommand{\erf}{\mathrm{erf}} % error function
\begin{document}

\preprint{AIP/123-QED}

\title{Computation of the solid-liquid interfacial free energy in hard spheres by means of thermodynamic integration}

\author{M. B\"ultmann}
\email{moritz.bueltmann@physik.uni-freiburg.de}
\affiliation{Physikalisches Institut, Albert-Ludwigs-Universität, 79104 Freiburg, Germany}
\author{T. Schilling}%
\email{tanja.schilling@physik.uni-freiburg.de}
\affiliation{Physikalisches Institut, Albert-Ludwigs-Universität, 79104 Freiburg, Germany}
\date{\today}

\begin{abstract}
We used a thermodynamic integration scheme, which is specifically designed for disordered systems, to compute the interfacial free energy of the solid-liquid interface in the hard-sphere model. We separated the bulk contribution to the total free energy from the interface contribution, performed a finite-size scaling analysis and obtained for the (100)-interface $\gamma=0.591(11)\boltzmann T\sigma^{-2}$.
\end{abstract}

\maketitle

\section{Introduction \label{sec:introduction}}

Monte Carlo simulation is widely used to compute the thermal equilibrium properties of atomistic materials models, such as e.g.~the interfacial free energy between different phases of a given substance\cite{frenkel2001understanding}. To compute the tension, the free energy and the stiffness of the interface between the solid and the liquid phase, various methods have been introduced over the past 50 years. The capillary-wave method was first introduced by Hoyt et al.~\cite{hoyt2001method} for metals. Morris and Song used it for the Lennard-Jones model~\cite{morris2003anisotropic} and Benet et al.~for the TIP4P/2005 model of water~\cite{benet2014study}. A method using thermodynamic integration techniques, called the \emph{cleaving method}, was first used by Broughton and Gilmer~\cite{broughton1986molecular} for the Lennard-Jones model. Davidchack and Laird refined this technique and used it for the hard-sphere model~\cite{davidchack2000direct,davidchack2010hard}. Between different methods, systematic errors due to finite size effects can differ. Hence the values predicted for a given system often do not agree, even if the simulations have been carried out with high precision.

Hard spheres are often used as a model to test simulation methods in statistical physics. Arguably, there are no hard-sphere-like atoms in nature, however
the model
captures the local ordering of atoms in the dense phases and has therefore been studied quite extensively in theory and colloid experiments \cite{alder1957,mulero2008,palberg2014,turci2014,royall2013}. Hard spheres are also interesting, because their phase behavior is athermal and the interfacial free energy is determined solely by entropy. The hard sphere interfacial free energy has been computed by means of various methods the past twenty years, partly producing contradictory results due to differences in systematic errors, and a final statement is still missing \cite{ohnesorge1994density,marr1994interfacial,davidchack2006anisotropic,davidchack2010hard,hartel2012tension,fernandez2012equilibrium,benjamin2015crystal}. In the following we would like to introduce a direct thermodynamic integration method and discuss the value we obtain with this method for the hard sphere system.

\section{Method \label{sec:theory}}
\subsection{Thermodynamic Integration \label{sec:theory:thermodynamicintegration}}

Thermodynamic integration is a method to compute differences in thermodynamic potentials. Consider the case in which we would like to compute the difference in free energy between a system of interest with a Hamiltonian $\mathcal{H}_{\mathrm{int}}$ and a reference system with a Hamiltonian $\mathcal{H}_{\mathrm{ref}}$, for which we can evaluate the free energy exactly. Further, assume that both Hamiltonians are defined on the same state space.
If we blend continuously from one Hamiltonian to the other by means of a combined Hamiltonian $\mathcal{H}(\varepsilon) = \varepsilon \mathcal{H}_{\mathrm{int}} + ({\varepsilon_{1}}-\varepsilon)\mathcal{H}_{\mathrm{ref}}$, where the ``switching'' parameter $\varepsilon$ is a real number, we obtain 
\begin{equation}
  \label{eq:TI}
F_{\mathrm{int}} - F_{\mathrm{ref}} := \varDelta F=
\int_{0}^{\varepsilon_{1}}\mathrm{d}\varepsilon^{\prime}\ \left\langle
\frac{\del\mathcal{H}}{\del\varepsilon}
\right\rangle_{N,V,T,\varepsilon^{\prime}}
\end{equation}
The angular brackets indicate the average taken with respect to the canonical ensemble for a given value of $\varepsilon$.
(We used a linear blending function here for simplicity, but it is straight-forward to implement other functional forms of $\mathcal{H}(\varepsilon)$ in order to optimize the performance of the method, see e.g.~refs.~\cite{steinbrecher2007,berryman2013free}.) The integrand $\left\langle
\frac{\del\mathcal{H}}{\del\varepsilon}
\right\rangle_{N,V,T,\varepsilon^{\prime}}$ can be computed by means of Monte Carlo sampling.

Thermodynamic integration requires a reference model which can be reached along a path that does not cross a first order phase transition. 
To construct an analytically solvable reference model for dense, disordered systems, we follow here the method introduced by Schmid and Schilling \cite{schilling2009computing,schmid2010method}: We construct a reference configuration of particles $\{\vec{r}_i^{\,\mathrm{ref}} | i = 1,\dots,N\}$ using as reference coordinates the particle positions of an arbitrary equilibrated configuration. Analogously to the Einstein crystal method\cite{frenkel1984}, a set of attractive wells $\varphi_i(\vec{r}_i-\vec{r}_i^{\,\mathrm{ref}})$, each of which only interacts with one particle $i$, is placed at each coordinate $\vec{r}_i^{\,\mathrm{ref}}$. Here we will use the same function for all wells and thus drop the index $i$ from $\varphi_i$. As the method is intended to study liquids, we need to take into account the possiblity that a particle moves infinitely far away from its reference position. Thus, in contrast to the Einstein crystal method, the potential $\varphi$ needs to be cut off at a finite value to prevent the sampling of a diverging function.
We introduce a cutoff radius $r_c$, above which the potential is zero, via $x=\left|\vec{r}_i-\vec{r}_i^{\,\mathrm{ref}}\right|/r_{\mathrm{c}}$.

For a linear reference potential
\begin{equation}
\varphi(x) =
\begin{cases}
0 & \mbox{for }x\geq1\\ 
x-1 & \mbox{for }x < 1\\
\end{cases}
\end{equation}
the  Helmholtz free energy can be obtained via integration by parts and using the Stirling approximation.
\begin{equation}
\begin{split}
&F_{\mathrm{ref}}(\varepsilon_{1})
\approx N\bigg[
\ln\left(\frac{N}{V}\right)\\
&-\ln\left(1+\frac{6V_c}{V}\frac{1}{\varepsilon^3_{1}}
\left(e^{\varepsilon_{1}}-1-\varepsilon_{1}
-\frac{\varepsilon^2_{1}}{2}
-\frac{\varepsilon^3_{1}}{6}
\right)\right)-1\bigg]
\end{split}
\end{equation}
where $V_{\mathrm{c}}$ is the volume of a sphere with radius $r_{\mathrm{c}}$.

Table~\ref{table:freeenergies} shows a list of other possible functional forms for $\varphi(x)$ and the free energies of the corresponding reference systems. We observed that equilibration times are shortest when using the linear well. However, if one uses molecular dynamics simulations instead of Metropolis Monte Carlo, potentials will be required that are differentiable in every point in space \cite{berryman2012free}. Then the functional forms listed in Table~\ref{table:freeenergies} can be useful.\\ 

\begin{table*}
\begin{ruledtabular}
\begin{tabular}{ll}
form of the potential well
& free energy of the corresponding $N$-particle system
$F(\varepsilon_{1})$
\\ \hline
\(\displaystyle
\varphi(x) =
\begin{cases}
0 & \mbox{for }x\geq1\\ 
x^2-1 & \mbox{for }x < 1\\
\end{cases} 
\)
&
\makecell[l]{
  \(\displaystyle
    N\left[\ln\left(\frac{N}{V}\right)-\right.
  \)
  \(\displaystyle
    \left.\ln\left(1+\frac{3V_c}{4V}\frac{1}{\sqrt{\varepsilon_{1}^3}}
    \left(\sqrt{\pi}e^{\varepsilon_{1}}\erf(\sqrt{\pi})
    -2\sqrt{\varepsilon}
    -\frac{4}{3}\sqrt{\varepsilon_{1}^3}
    \right)\right)-1\right]
  \)
}
\\
\(\displaystyle
\varphi(x) =
\begin{cases}
0 & \mbox{for }x\geq1\\ 
x^3-1 & \mbox{for }x < 1\\
\end{cases} 
\)
&
\(\displaystyle
N\left[
\ln\left(\frac{N}{V}\right)
-\ln\left(1+\frac{V_c}{V}\frac{1}{\varepsilon_{1}}
\left(e^{\varepsilon_{1}}-1-\varepsilon_{1}
\right)\right)-1\right]
\)
\\
\(\displaystyle
\varphi(x) =
\begin{cases}
0 & \mbox{for }x\geq1\\
\sqrt{x}-1 & \mbox{for }x < 1\\
\end{cases} 
\)
&
\(\displaystyle
N\left[
\ln\left(\frac{N}{V}\right)
-\ln\left(1+\frac{720V_c}{V}\frac{1}{\varepsilon_{1}^6}
\left(e^{\varepsilon_{1}}
-\sum_{k = 0}^{6}\frac{\varepsilon_{1}^k}{k!}
\right)\right)-1\right]
\)
\\
\(\displaystyle
\varphi(x) =
\begin{cases}
0 & \mbox{for }x\geq1\\ 
x^{\frac{3}{n}}-1 & \mbox{for }x < 1\\
\end{cases} 
\)
&
\makecell[l]{
  \(\displaystyle
    N\left[
    \ln\left(\frac{N}{V}\right)
    -\ln\left(1+\frac{n!V_c}{V}\frac{1}{\varepsilon_{1}^n}
    \left(
    \sum_{k = n+1}^{\infty}\frac{\varepsilon_{1}^k}{k!}
    \right)\right)-1\right]
  \)
for $n\in\mathbb{N}\setminus\{0\}$
}
\\
\end{tabular}
\end{ruledtabular}
\caption{Free energy expressions for model systems with different well potentials with finite range in three dimensions. The function $\erf(\dots)$ denotes the error function. The last expression is a generalization of the two expressions preceding it.\label{table:freeenergies}}
\end{table*}

In the liquid-solid coexistence regime, the high density renders equilibriation and decorrelation difficult, because particles may be blocked from moving into their wells for many Monte Carlo steps. To circumvent this problem we used a \emph{swap move} as introduced in ref.~\cite{schilling2009computing,schmid2010method}.

Finally, we need to take into account one specificity of the hard sphere model. As the reference Hamiltonian does not contain pair potentials, all pair interactions need to be switched off when the parameter $\varepsilon$ approaches the value $\varepsilon_1$. However, the hard sphere interaction potential diverges for overlapping spheres, while we can only use a set of finite values for $\varepsilon$ to evaluate the ensemble averages in eq.~\ref{eq:TI}. To circumvent this problem, we used a finite-valued repulsive potential $V_{\rm sph}$ between the spheres  - finite, but large enough for small $\varepsilon$ to ensure that the probability of two particles overlapping was negligible
\begin{equation}
V_{\rm sph}(\vec{r}_i,\vec{r}_j,\varepsilon)=
\begin{cases}
A \left(1-\frac{\varepsilon}{\varepsilon_{1}}\right)^{B} & \mbox{for } |\vec{r}_i-\vec{r}_j| < \sigma\\
0 & \mbox{for } |\vec{r}_i-\vec{r}_j| \geq \sigma\\
\end{cases}
\end{equation}
where $\sigma$ is the diameter of the hard spheres. We set $A=40\,\boltzmann T$.
To optimize the equilibration times for all $\varepsilon$, we used a polynomial of order $B=4$ to switch off this pair potential.

In summary, the free energy difference between the hard sphere system and the reference system then has the form

\begin{equation}
\begin{split}
\varDelta F=
\int_{\varepsilon^{\prime} = 0}^{\varepsilon^{\prime}=\varepsilon_{1}}
\mathrm{d}\varepsilon^{\prime}\ \bigg\langle&
-\frac{N_{\mathrm{overlaps}}AB}{\varepsilon_{1}}
\left(
1-\frac{\varepsilon}{\varepsilon_{1}}
\right)^{B-1}\\
&+\sum_{i = 1}^{N}\varphi\left(
\frac{\left|\vec{r}_i-\vec{r}_i^{\,\mathrm{ref}}\right|}{r_{\mathrm{c}}}
\right)
\bigg\rangle_{N,V,T,\varepsilon^{\prime}}
\end{split}
\end{equation}

We show in detail in section \ref{subs:ThermoResults} how this expression can be used to compute the interfacial free energy.

\subsection{Pressure Tensor}
To check whether the simulated system was subject to mechanical stress, we computed the local excess pressure tensor. For hard spheres Allen showed\cite{allen2006evaluation}, that the following limit holds:

\begin{equation}\label{eqn:pressureTensor}
\frac{\mathrm{P}^{\mathrm{ex}}_{\alpha\beta}}{\boltzmann T} 
=\lim_{\xi\rightarrow 0^{+}}\frac{1}{V\xi}\left\langle
\sum_{i < j}^{N}\phi_{ij}
\frac{\left(\vec{r}_{ij}\right)_{\alpha}\left(\vec{r}_{ij}\right)_{\beta}}
{|\vec{r}_{ij}|^{2}}
\right\rangle
\end{equation}

The double sum is taken over all unique particle pairs. The $\alpha$-th component of the distance vector $\vec{r}_{ij}$ between the particle pair $i,j$ is given by $\left(\vec{r}_{ij}\right)_{\alpha}$. $\phi_{ij}$ is a function that is either $1$, if the particle pair $i,j$ is overlapping or $0$ otherwise. The brackets $\langle\cdot\rangle$ denote the thermodynamic ensemble average. Hence, to compute the pressure tensor approximately, one increases the hard sphere diameter $\sigma$ by a factor $1+\xi$ with $\xi\ll1$ and counts the hard sphere overlaps that occur.

\section{Simulations \label{sec:simulations}}
\subsection{Setup}
We carried out Metropolis Monte Carlo simulations with systems of different geometry and size, in cuboid simulation boxes with periodic boundary conditions.
The systems consisted of $N=1\,097\dots38\,993$ particles. The number density was $\rho=N/V=0.991\sigma^{-3}$. The geometries of the systems could be divided into two classes. One class contained the systems with constant shorter dimensions $L_x = L_y \equiv L = 9.3978\sigma$ and a varying longer dimension $L_z = 12.5304\sigma\dots125.304\sigma$. The other class contained the systems with constant longer dimension $L_z = 62.652\sigma$  and varying shorter dimensions $L = 6.2652\sigma\dots25.0608\sigma$. We chose these geometries such that the $z$-dimension was always significantly larger than the other two. This fixes the solid-liquid interface parallel to the $x$-$y$ plane. Thus its projected area is $L^2$.

To obtain well equilibrated systems in coexistence, two smaller systems -- one solid, the other liquid -- were merged to form a larger system. The solid part was set up as an fcc-crystal with the equilibrium density of the solid at coexistence ($\rho_{\text{\tiny solid}}^{\text{\tiny coex}}=1.0408\sigma^{-3}$) reported in \cite{hoover1968melting}. Hence the unit cell dimensions of the crystal were $1.5663\sigma$ and the simulation box dimensions are integral multiples of it. The interface orientation of the crystal was $(100)$. The liquid part was set up in a box of the same size, but not in a liquid state. Rather, it was set up in a crystal structure with the equilibrium density of the liquid at coexistence. First, the solid particles were kept fixed in their places and the liquid particles were equilibrated by performing $2N\cdot10^6$ Monte Carlo (MC) steps. Some of the liquid particles crystallized on the fixed solid interface. Afterwards the whole system was simulated for another $2\cdot10^6$ MC sweeps ($2N\cdot10^6$ steps) to reach an equilibrium coexistence state.

To distinguish between solid and liquid particles, we used the bond order parameter described in ref.~\cite{steinhardt1983bond}.

\subsection{Thermodynamic Integration}
\label{subs:ThermoResults}
\begin{figure}[!b]
\centering
\includegraphics[width=\linewidth]{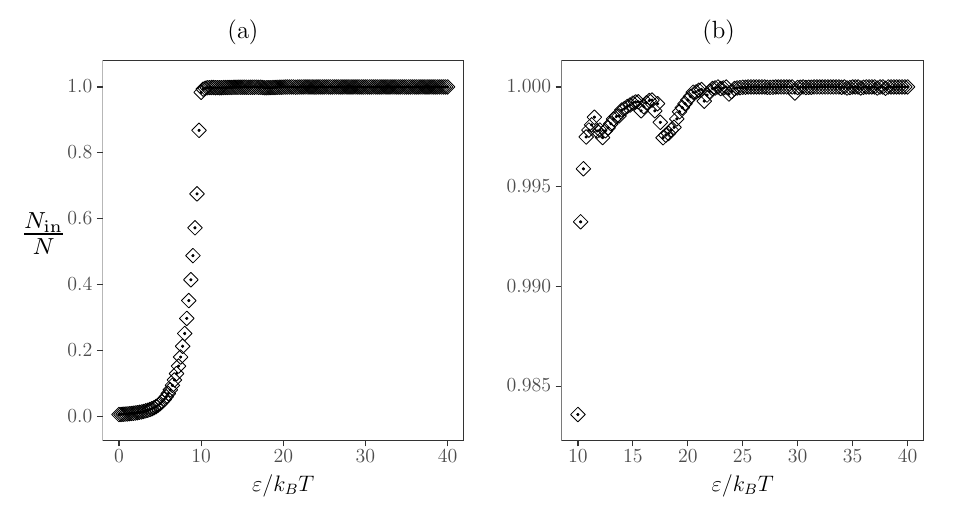}
\caption{Ratio of particles $N_{\mathrm{in}}/N$ that are in the attraction range of their respective potential wells for the system of size $L = 25.0608\sigma$, $L_z = 62.652\sigma$, $N = 38,993$. Note that graph (b) is an enlarged version of graph (a) which shows all $\varepsilon$-values for which the ratio is close to $1$.\label{fig:proportioninwells}}
\end{figure}

The thermodynamic integration procedure was applied to all equilibrated systems in phase coexistence. We used one particle configuration for each system size as a reference configuration, which determined the coordinates of the potential well centers. The total free energy is given by 
\begin{eqnarray}\label{eqn:freeenergyfinal}
F(\varepsilon_{0} = 0)
&=& N\left[
\ln\left(\frac{N}{V}\right)-1 \right.\\
&-&\left. \ln\left(1+\frac{6V_c}{V}\frac{1}{\varepsilon^3_{1}}
\left(e^{\varepsilon_{1}}-1-\varepsilon_{1}
-\frac{\varepsilon^2_{1}}{2}
-\frac{\varepsilon^3_{1}}{6}
\right)\right)\right]\nonumber\\
&+&
\int_{\varepsilon^{\prime} = 0}^{\varepsilon^{\prime}
= \varepsilon_{1}}\mathrm{d}\varepsilon^{\prime}
\ \left\langle
-\frac{N_{\mathrm{overlaps}}AB}{\varepsilon_{1}}
\left(
1-\frac{\varepsilon}{\varepsilon_{1}}
\right)^{B-1} \right.\nonumber\\
&+& \left. \sum_{i = 1}^{N}\varphi\left(
\frac{\left|\vec{r}_i-\vec{r}_i^{\,\mathrm{ref}}\right|}{r_{\mathrm{c}}}
\right)
\right\rangle_{N,V,T,\varepsilon^{\prime}}\nonumber
\hspace{1em}\mbox{,}
\end{eqnarray}
where the ensemble average term in angular brackets needs to be determined by simulation.
The parameter $\varepsilon$ that switches between the model system and the hard sphere system Hamiltonian was chosen to be in the range $\varepsilon\in[0,40]\ \boltzmann T$. One reason for this choice of the range is that test runs showed that $99\%$ of particles find their respective wells at around $\varepsilon=12\boltzmann T$, and that on average fewer than $0.1\%$ of all particles are outside of their wells at about $\varepsilon=30\boltzmann T$. In figure~\ref{fig:proportioninwells} we show the dependence of the number of particles inside the potential wells on $\varepsilon$ for the system size $L = 25.0608\sigma$, $L_z = 62.652\sigma$, $N = 38,993$. (These numbers change slightly with system size, but they are close enough to allow us to use the same integration interval for all systems.) 
The integration range was sampled at $161$ evenly distributed points (abscissas of the integral) which is, as we will see later, the main source of error of the method. We chose the cutoff radius to be $r_c=2\sigma$, because the number of sweeps necessary to equilibrate the systems is minimized for this value.

To compute the integral in eqn.~\eqref{eqn:freeenergyfinal}, for every abscissa an average value of $\del\mathcal{H}/\del\varepsilon$ is required. Hence, after an equilibration period of $3N\cdot10^5$ Monte Carlo steps, between $1\cdot10^3$ and $2\cdot10^4$ samples of this quantity were recorded (depending on the system size) with $200$ decorrelation sweeps between each pair of samples. This is done separately for each abscissa. The thermodynamic integration process was done forwards (starting at $\varepsilon=0\boltzmann T$) and backwards (starting at $\varepsilon=40\boltzmann T$). Since no hysteresis occured , there is no first order phase transition present. As the quantity $\langle\del\mathcal{H}/\del\varepsilon\rangle$ is only obtained at a finite number of abscissas with a finite accuracy, the integral needs to be estimated numerically.

We used three different quadrature rules to approximate the integral: the trapezoidal rule, Simpson's rule and Romberg-integration with Richardson-extrapolation. 
The error of the integral due to the uncertainty of the data points $\langle\del\mathcal{H}/\del\varepsilon\rangle$ was estimated with a parametric bootstrapping method. For that we assumed that every data point stems from a Gaussian distribution. From these distributions random numbers were generated that were then used as \emph{artificial} data sets for the integration scheme instead of the real data. From the obtained integral values of the \emph{artificial} data sets one can then estimate an error for the integral, and hence, for the free energy. However, the errors of this kind produce a relatively small error in the free energy (about $3\cdot10^{-3}\%$) and they will therefore be neglected. The numerical error due to the finite number of $161$ abscissas is the major contribution to the total error.\\

\begin{figure}
\centering
\includegraphics[width=\linewidth]{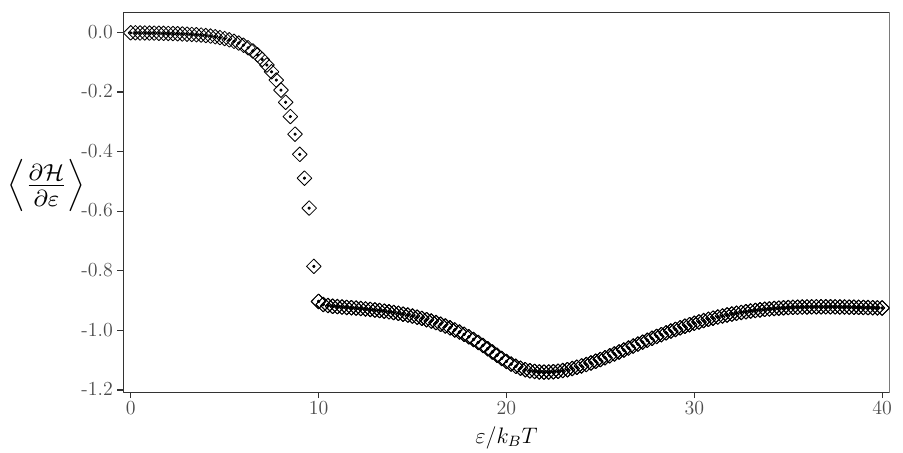}
\caption{Integrand used for thermodynamic integration of a system of size $L = 9.3978\sigma$, $L_z = 62.652\sigma$, $N = 5483$. $\langle\del\mathcal{H}/\del\varepsilon\rangle$ has been divided by the number of particles, to allow for comparison with other systems. Each data point is an average over $9000$ independent samples. The error bars cannot be seen, because they are smaller than the center dots of the symbols.\label{fig:tdiCurve}}
\end{figure}

Fig.~\ref{fig:tdiCurve} shows a thermodynamic integration curve for a system of size $L = 9.3978\sigma$, $L_z = 62.652\sigma$. The error bars are not visible, because they are smaller than the center dots of the diamonds. (Note that in the graph, $\langle\del\mathcal{H}/\del\varepsilon\rangle$ has been divided by the number of particles to allow for comparison with other systems.)

\begin{figure}
\centering
\includegraphics[width = \linewidth]{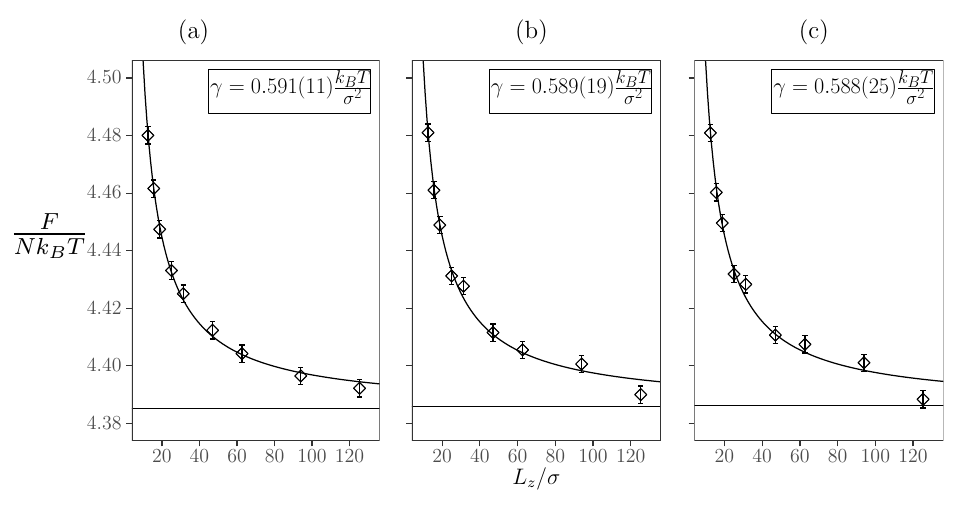}
\caption{Free energy per particle as a function of the length $L_z$ for different integration schemes, namely: (a) trapezoidal rule, (b) Simpson's rule, (c) Romberg integration. The interfacial area $L^2$ is the same in all systems ($L = 9.3978\sigma$). The curved line is a non-linear least squares fit of the free energies to obtain the interfacial free energy $\gamma$. The straight line represents the bulk free energy value obtained from the fit.\label{fig:freeenergies:constantInterface}}
\end{figure}

\begin{figure}
\centering
\includegraphics[width = \linewidth]{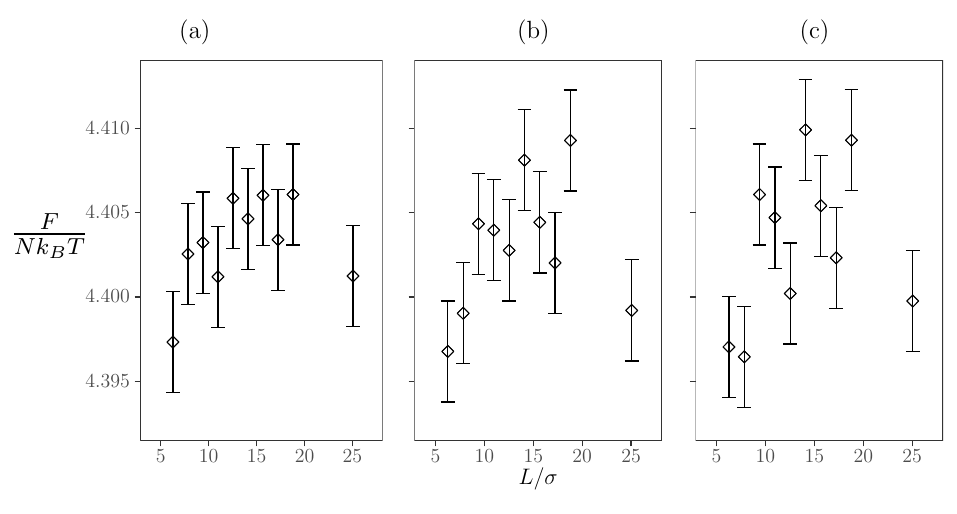}
\caption{Free energy per particle as a function of the interfacial edge length $L$ for different integration schemes, namely: (a) trapezoidal rule, (b) Simpson's rule, (c) Romberg integration. The longer dimension $L_z$ is the same for all systems ($L_z = 62.652\sigma$)\label{fig:freeenergies:constantLength}}
\end{figure}

The free energy per particle $f\equiv F/N$ as a function of the system size and integration scheme is shown in figure~\ref{fig:freeenergies:constantInterface}  for varying $L_z$ and in figure~\ref{fig:freeenergies:constantLength} for varying $L$. The error of the free energies $s_{F/N} = 0.003\boltzmann T$ was estimated by using $401$ abscissas for the three smallest systems and comparing the results to the free energies obtained with $161$ abscissas.

For an infinitely long system $L_z\rightarrow\infty$ the contribution to the free energy of the two interfaces $f_{\mathrm{interface}}$ which are not varying in size $L$ vanishes. The remaining free energy per particle should thus be equal to the average bulk free energy per particle $f_{\mathrm{bulk}}$. Moreover the interfacial contribution to the free energy per particle should be proportional to the inverse length of the system $1/L_z$. The proportionality hence contains $\gamma$ as follows

\begin{equation}\label{eqn:freeEnergyModel}
\begin{split}
f(L_{z}):=\frac{F(L_z)}{N(L_z)} &= f_{\mathrm{bulk}} + f_{\mathrm{interface}}(L_z)\\
&= f_{\mathrm{bulk}} + \frac{2\gamma L^2}{N(L_z)}\\
&= f_{\mathrm{bulk}} + \frac{2\gamma}{\rho L_z}\\
\end{split}
\end{equation}

However, this expression does not yet account for systematic errors due to finite-size effects. Schmitz et al.~identified three finite-size contributions to $\gamma$ by phenomenological considerations \cite{schmitz2014logarithmic}

\begin{equation}
\gamma = \gamma_{\infty} - P\frac{\ln(L_z)}{L^2} + Q\frac{\ln(L)}{L^2} + R\frac{1}{L^2}
\end{equation}

where $P\geq0$, $Q\geq0$ and $R$ are constants. $\gamma_{\infty}$ is the interfacial free energy for the system with infinite size. $P$ and $Q$ only depend on the dimension of the system, on the statistical ensemble and on whether or not periodic boundary conditions are employed. In our case the constants are $P=3/4$ and $Q=1/2$ ($3$ dimensions, periodic boundary conditions, canonical ensemble). The constant $R=0.95(37)$ needs to be estimated and can be extracted from \cite{benjamin2015crystal}, where the finite size scaling of the interfacial free energy was investigated. After incorporating the finite size scaling into our fit model~\eqref{eqn:freeEnergyModel} it has the following form\\
\begin{equation}
f(L_{z}) = f_{\mathrm{bulk}} + \frac{2}{\rho}\left(\left(\gamma_{\infty}+C\right)\frac{1}{L_z}
-\frac{P}{L^2}\frac{\ln(L_z)}{L_z}
\right)
\end{equation}

where $C:=Q\ln(L)/L^2+R/L^2$ is a known constant. This expression can be fitted to the free energies per particle obtained from systems with the same $L$ and varying $L_z$. For the fitting, we used the Levenberg-Marquardt algorithm provided by the \textbf{R}-package 'minpack.lm'~\cite{elzhov2016minpack} to find the minimum of the sum of weighted least squares in parameter space. The fit curves are also shown fig.~\ref{fig:freeenergies:constantInterface}. The resulting interfacial free energy values are


\begin{equation}
  \label{resultsThermo}
\begin{split}
\gamma_{\mathrm{trap}}&=0.591(11)\frac{\boltzmann T}{\sigma^2}\\
\gamma_{\mathrm{simp}}&=0.589(20)\frac{\boltzmann T}{\sigma^2}\\
\gamma_{\mathrm{romb}}&=0.588(25)\frac{\boltzmann T}{\sigma^2}
\end{split}
\hspace{1em}\mbox{.}
\end{equation}

The three different values $\gamma_{\mathrm{trap}}$,$\gamma_{\mathrm{simp}}$,$\gamma_{\mathrm{romb}}$ stem from the different integration methods used to perform the free energy calculation.
The relative errors can be reduced by improving the accuracy of the numerical quadrature and the number of simulated systems.

As we expect a constant relation between the interfacial area $L^2$ and the free energy per particle, it is not possible to extract the interfacial free energy from fig.~\ref{fig:freeenergies:constantLength} directly. However, if one knows the free energy of the solid and liquid bulk phases at their respective coexistence densities this is still possible. We determined the densities far away from the interface and set up separate simulations to determine the bulk free energies at these densities. Since our crystal structure was set up with a fixed density of $1.0408\sigma^{-3}$ our resulting values are not the true coexistence densities. Our values do compare well to~~\cite{hoover1968melting} $\rho_l=0.943(4)\sigma^{-3}, \rho_s=1.041(4)\sigma^{-3}$, but turn out to by slightly larger than~~\cite{frenkel2001understanding} with $\rho_l=0.9391\sigma^{-3}, \rho_s=1.0376\sigma^{-3}$,~~\cite{de2008estimating} with $\rho_l=0.935(2)\sigma^{-3}, \rho_s=1.033(3)\sigma^{-3}$, and~~\cite{noya2008determination} with $\rho_l=0.9375(14)\sigma^{-3}, \rho_s=1.0369(33)\sigma^{-3}$.

\begin{equation}\label{densities}
\begin{split}
\rho_{\mathrm{liquid}}^{\mathrm{coex}}&=0.9391(10)\frac{1}{\sigma^{3}}\\
\frac{F_{\mathrm{liquid}}}{N}&=3.745(3)\boltzmann T\\
\frac{F_{\mathrm{solid}}}{N}&=4.953(3)\boltzmann T
\end{split}
\hspace{1em}\mbox{,}
\end{equation}

The strategy is then to subtract the bulk free energies $F_{\mathrm{liquid}}$ and $F_{\mathrm{solid}}$  weighted by the particle number in the respective phase from the free energy of the systems in coexistence to be left with the total interfacial free energy. (This approach is similar to interfacial free energy calculations at hard walls, where the free energy difference between a system with and without hard walls is calculated~\cite{deb2012methods}.) However, the bond order parameter analysis did not allow for a sufficiently precise determination of the particle numbers in the two phases to produce a value for $\gamma$ that is as accurate as eqns.~\ref{resultsThermo}.

\subsection{Pressure Tensor Analysis}\label{sec:simulations:pressure}

In those systems which contain two phases at coexistence, we expect the pressure to be inhomogeneous  in the $x$- and $y$-direction parallel to the interface and homogeneous in the $z$-direction normal to the interface. Thus computing the spacial profile of the pressure tensor in the $z$-direction can be helpful in detecting systems that are out of thermal equilibrium.

Due to the periodic boundary conditions, the system as a whole can perform translations in the box without a free energy cost. We therefore needed to center the system before we computeed the pressure tensor. Using the bond-order parameter, we mapped every particle to a phase. We then computed the centers of mass
of the phases and translated every particle by an amount that put the center of mass of the solid phase into the center of the box.

\begin{figure}
\includegraphics[width = \linewidth]{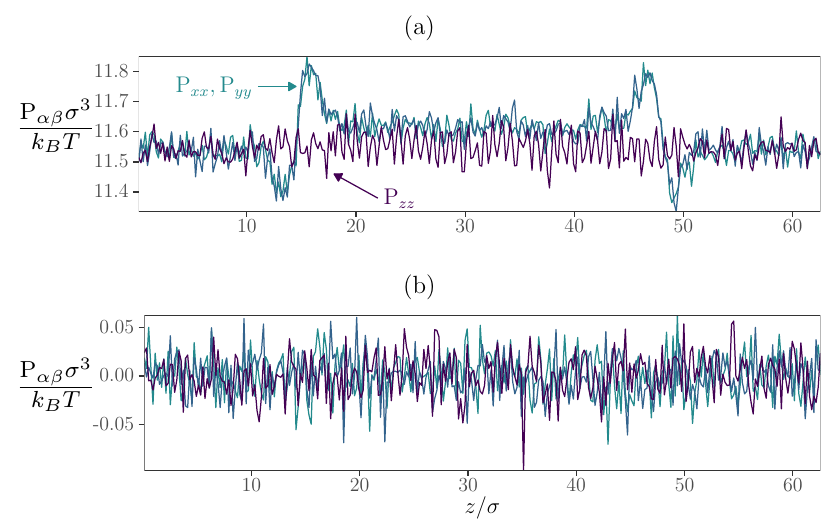}
\caption{Pressure tensor profile for $L_z = 62.652\sigma$ and $L^2 = (9.3978\sigma)^2$. Note that the ideal gas contribution has been added. The errors of the diagonal elements (a) and the off-diagonal elements (b) are respectively $0.035\boltzmann T\sigma^{-3}$ and $0.025\boltzmann T\sigma^{-3}$.\label{fig:pressureTensor}}
\end{figure}

The pressure tensor was computed for all systems in coexistence. The bin width was chosen to be as small as possible, without making the statistical error too large to conclude whether the systems was free of stresses ($\varDelta z = 0.2\sigma$). Fig.~\ref{fig:pressureTensor} shows the pressure tensor for the system with $L = 9.3978\sigma, L_z = 62.652\sigma$. The scaling factor was chosen close enough to $\xi=0$ such that the systematic effects only play a minor role, but large enough that statistics allow to resolve potential stresses ($\xi = 3\cdot10^{-4}$). The measured values are an average over $2\cdot10^5$\dots$7\cdot10^6$ samples (depending on the system size) with $200$ sweeps between each sample.

The normal pressure value $P_{zz}$ is, as expected, homogeneous. The tangential components $P_{xx}$ and $P_{yy}$ do not decay entirely to the bulk value in the crystal (deviation of $<1\%$), which indicates a small but negligible stress.

In the liquid part all diagonal elements approach the same value in regions far from the interface, which means that bulk behavior is recovered. All off-diagonal elements are compatible with zero within their margin of error (as can be seen in the lower panel of figure \ref{fig:pressureTensor} for one example).

With eqn.~\eqref{eqn:pressureTensor} we can also compute the coexistence pressure of the systems by averaging the diagonal elements $P_{\text{coex}}=(P_{xx}+P_{yy}+P_{zz})/3$. Using a finite $\xi$ leads to a variance-bias trade-off problem. For smaller values of $\xi$ the probability that two spheres overlap is relatively small which leads to bigger statistical fluctuations. However, increasing $\xi$ to reduce the statistical error leads to a systematic error, because the assumptions that were made to derive eqn.~\eqref{eqn:pressureTensor} do not hold anymore. To obtain the best possible value, we performed the calculations for different $\xi=1\cdot10^{-2}, 2\cdot10^{-3}, 4\cdot10^{-4}, 8\cdot10^{-5}$. With these data it is possible to extrapolate to $\xi=0$ by means of linear regression. Thus, we can obtain a fairly good approximation of the coexistence pressure.
(Note that the ideal gas pressure ($P^{\mathrm{id}}=\rho\boltzmann T$) was added to all measured pressure values to obtain the total pressure.)
\begin{figure}
\centering
\includegraphics[width = \linewidth]{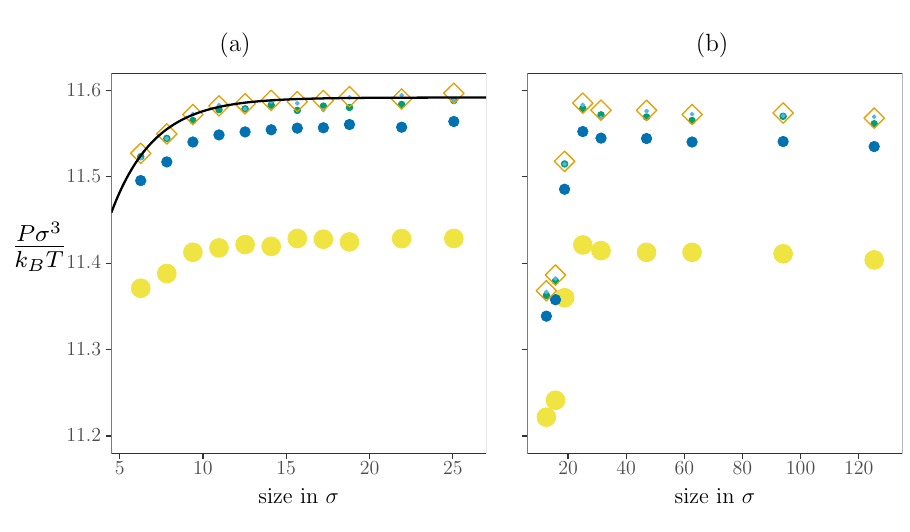}
\caption{Total pressures for different scaling factors (different point sizes, different colors online) and different system sizes. For every system a linear regression was performed to extrapolate to a scaling factor of $0$ (diamonds). The pressure $P_{\mathrm{coex}}$ was calculated with an exponential fit (black line). In graph (a) systems have the same length $L_z=62.652\sigma$ but vary in their interface dimensions $L$. In graph (b) the systems have the same interface size $L=9.398\sigma$ but vary in length $L_z$.\label{fig:pressureScaling}}
\end{figure}

The extrapolations for different system sizes and the finite size scaling are shown in figure~\ref{fig:pressureScaling}.
The systems on the right-hand side have a fairly small interface size $L=9.398\sigma$.
Thus their pressure values are affected by both finite size scalings.
However, the qualitative behavior shows that for systems of length $L_z=62.652\sigma$ the finite size scaling in $z$-direction is negligible.
Thus, we can extract the coexistence pressure from the graph on the left-hand side, where all systems have this length.
By using an exponential fit, we can extrapolate the coexistence pressure to infinitely large interface sizes $L$.
Depending on the system size the pressure was computed between $4\cdot10^4$ and $3\cdot10^6$ times with $200$ sweeps between each pair of samples. The error is estimated based on the variance of the fit parameters. In conclusion we obtain

\begin{equation}
P_{\text{coex}} = 11.591(10)\frac{\boltzmann T}{\sigma^3}
\end{equation}

This result agrees with many previous studies, but is is slightly larger -- however statistically more accurate -- than most of them (e.g. $P_{\text{coex}} = 11.5727(10)\frac{\boltzmann T}{\sigma^3}$\cite{fernandez2012equilibrium}, $P_{\text{coex}} = 11.57(10)\frac{\boltzmann T}{\sigma^3}$ \cite{fortini2006phase}, or $P_{\text{coex}} = 11.54(4)\frac{\boltzmann T}{\sigma^3}$ \cite{noya2008determination}).
We expect the true pressure to be slightly lower than our measurements, because the neglected finite size scaling in $z$-direction seems to decay very slowly in a subexponential manner.
In addition, on approach of $\xi=0$ the pressure value might increase more slowly than linearly, but the error bars are too big to be certain.

\section{Conclusion and Discussion}
We have computed the solid-liquid interfacial free energy in hard spheres by means of a thermodynamic integration with respect to a reference model, which can be solved exactly. Our results for the interfacial free energy of the (100)-interface are $\gamma=0.591(11)\boltzmann T\sigma^{-2}$, $\gamma=0.589(20)\boltzmann T\sigma^{-2}$ and $\gamma=0.588(25)\boltzmann T\sigma^{-2}$, depending on the integration scheme.
These values are lower than predictions by density functional theory, e.g.~$\gamma=0.664(2)\boltzmann T\sigma^{-2}$ \; \cite{oettel2010free}, and than simulation results obtained with the cleaving method, e.g.~$\gamma=0.62(2)\boltzmann T\sigma^{-2}$ \; \cite{davidchack2000direct}. However more recent studies using the cleaving method~\cite{davidchack2010hard} and \cite{benjamin2015crystal} produced a slightly lower value of $\gamma=0.5820(19)\boltzmann T\sigma^{-2}$ and $\gamma=0.596(2)\boltzmann T\sigma^{-2}$, which is in agreement with our results. Another thermodynamic integration method, called mold integration, produced similar results $\gamma=0.586(8)\boltzmann T\sigma^{-2}$ \; \cite{espinosa2014mold}. Capillary wave analysis yielded even lower values, as e.g.~$\gamma=0.56(2)\boltzmann T\sigma^{-2}$ from~\cite{davidchack2006anisotropic}.

\begin{acknowledgments}
  We thank M.~Allen for pointing out a mistake in the pressure tensor calculation. 
We acknowledge the support by the state of Baden\hspace{.2ex}-W\"urttemberg through bwHPC and the German Research Foundation (DFG) through grant no INST 39/963-1 FUGG (bwForCluster NEMO), RV bw17B003.
\end{acknowledgments}

\bibliography{paper.bib}  % Produces the bibliography via BibTeX.

\end{document}